\documentclass[a4paper,11pt]{article}
\usepackage{pos}
\usepackage[english]{babel}
\usepackage{amsmath}
\usepackage{float}
\usepackage{graphicx}
\usepackage{subcaption}
\usepackage{physics}
\usepackage{siunitx}
\usepackage{sidecap}
\usepackage{wrapfig}
\usepackage{dsfont}
\usepackage{color}
\usepackage{hyperref}

\renewcommand{\printHeadAuthors}{Torralba Paz et al.}

\title{Prediction and Anomaly Detection of accelerated particles in PIC simulations using neural networks}
 \ShortTitle{Machine learning methods for Particle-In-Cell simulations}

\author*[a]{Gabriel Torralba Paz}
\author[b, c]{Artem Bohdan}
\author[a]{Jacek Niemiec}

\affiliation[a]{Institute of Nuclear Physics Polish Academy of Sciences, PL-31342 Krakow, Poland}

\affiliation[b]{Max-Planck-Institut für Plasmaphysik, Boltzmannstr. 2, DE-85748 Garching, Germany}

\affiliation[c]{Excellence Cluster ORIGINS, Boltzmannstr. 2, DE-85748 Garching, Germany}

\emailAdd{gtorralba@ifj.edu.pl}

\abstract{Acceleration processes that occur in astrophysical plasmas produce cosmic rays that are observed on Earth. To study particle acceleration, fully-kinetic particle-in-cell (PIC) simulations are often used as they can unveil the microphysics of energization processes. Tracing of individual particles in PIC simulations is particularly useful in this regard. However, by-eye inspection of particle trajectories includes a high level of bias and uncertainty in pinpointing specific acceleration mechanisms that affect particles. Here we present a new approach that uses neural networks to aid individual particle data analysis. We demonstrate this approach on the test data that consists of 252,000 electrons which have been traced in a PIC simulation of a non-relativistic high Mach number perpendicular shock, in which we observe the two-stream electrostatic Buneman instability to pre-accelerate a portion of electrons to nonthermal energies. We perform classification, regression and anomaly detection by using a Convolutional Neural Network. We show that regardless of how noisy and imbalanced the datasets are, the regression and classification are able to predict the final energies of particles with high accuracy, whereas anomaly detection is able to discern between energetic and non-energetic particles. The methodology proposed may considerably simplify particle classification in large-scale PIC and also hybrid kinetic simulations.}

\ConferenceLogo{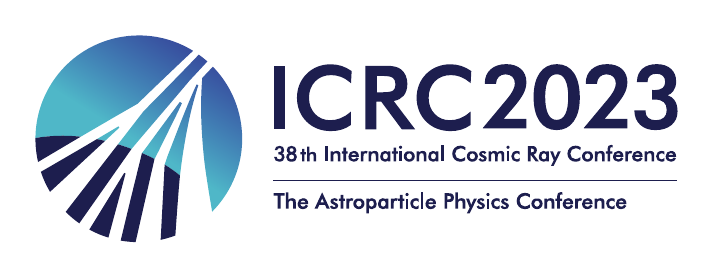}

\FullConference{
38th International Cosmic Ray Conference (ICRC2023)\\
  26 July - 3 August, 2023\\
  Nagoya, Japan}
  
\begin{document}
\maketitle

\section{Introduction}
Cosmic rays (CRs) are high-energy charged particles accelerated to nearly the speed of light carrying immense kinetic energies that can reach up to \SI{e21}{\electronvolt}. 
Such particles can be produced via a number of mechanisms, including the first- and second-order Fermi acceleration~\cite{Fermi1,Fermi2}, magnetic reconnection~\cite{mag_rec}, shock drift acceleration (SDA)~\cite{SDA} and stochastic SDA~\cite{SSDA} processes. To understand the microphysics of the acceleration processes, kinetic plasma simulation techniques, e.g., particle-in-cell (PIC) methods \cite{PIC}, are often employed. PIC simulations, which follow individual particle trajectories in self-generated electromagnetic fields, have recently facilitated significant progress in understanding complex dynamics and interactions within astrophysical plasma environments, including shock physics~\cite{Bohdan_Buneman}.

For an in-depth analysis of particle acceleration mechanisms, many of the PIC codes employ particle tracing. This feature enables the tracking of particles that are part of the plasma itself, as opposed to test-particle simulations \cite{test_par_sim}
However, disentangling acceleration processes requires a large amount of data, a thorough examination of the trajectories and an intuition to determine which particles were affected by the potential mechanisms involved. This could introduce a significant bias. 
Hence, we propose a novel approach
for analysing particle tracing data based on Neural Networks (NNs). This method enables fast and reliable post-processing of thousands of particle trajectories. 
The NN can effectively identify particles that have been energized by detecting the distinct patterns imprinted in momentum space during acceleration.
NN have been used before in various high-energy astrophysical applications, e.g., identification of neutron star mergers \cite{mergers}, image processing in Cherenkov telescopes \cite{gr_detector}, or anomaly detection in gravitational wave detectors \cite{anomalies_grav_waves}. In our
study, we use NN to select non-thermal particles from all-particle distribution obtained in kinetic plasma simulations. We utilise two algorithms, Regression and Anomaly Detection, to analyze traced particles. This approach is unique and has not been previously explored.

\section{Description of the particle dataset}
\begin{figure}
    \centering
    \includegraphics[width=0.99\linewidth]{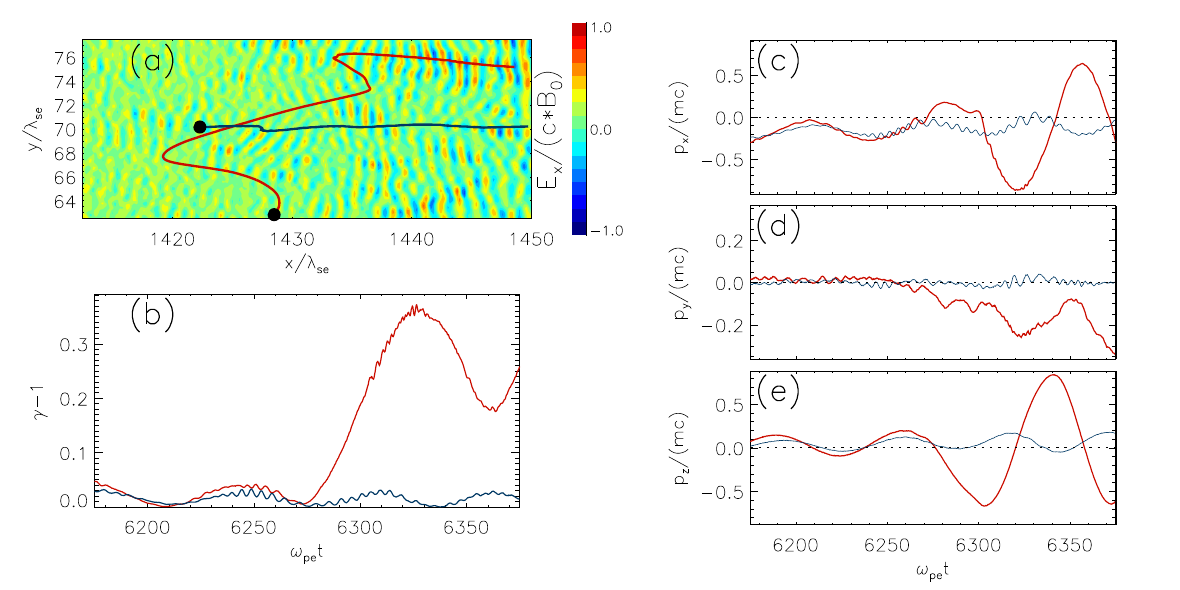}
    \caption{Time series for two sample electrons travelling through the Buneman instability region. Shown are particle trajectories overlaid on a map of the $E_x$ electric field (a), the evolution of particle Lorentz factor (b), and the evolution of particle momentum components (c-e). The electron data shown with a red (blue) line belongs to the non-thermal (thermal) population. Spatial coordinates are given in units of the electron skin depth, $\lambda_{\rm se=c/\omega_{\rm se}}$. Here $c$ is the speed of light and $\omega_{\rm pe}$ is the electron plasma frequency, which is also the unit of time.}
    \label{fig:bun_traj}
\end{figure}

\begin{SCfigure}
 \centering
\includegraphics[width=0.8\textwidth]{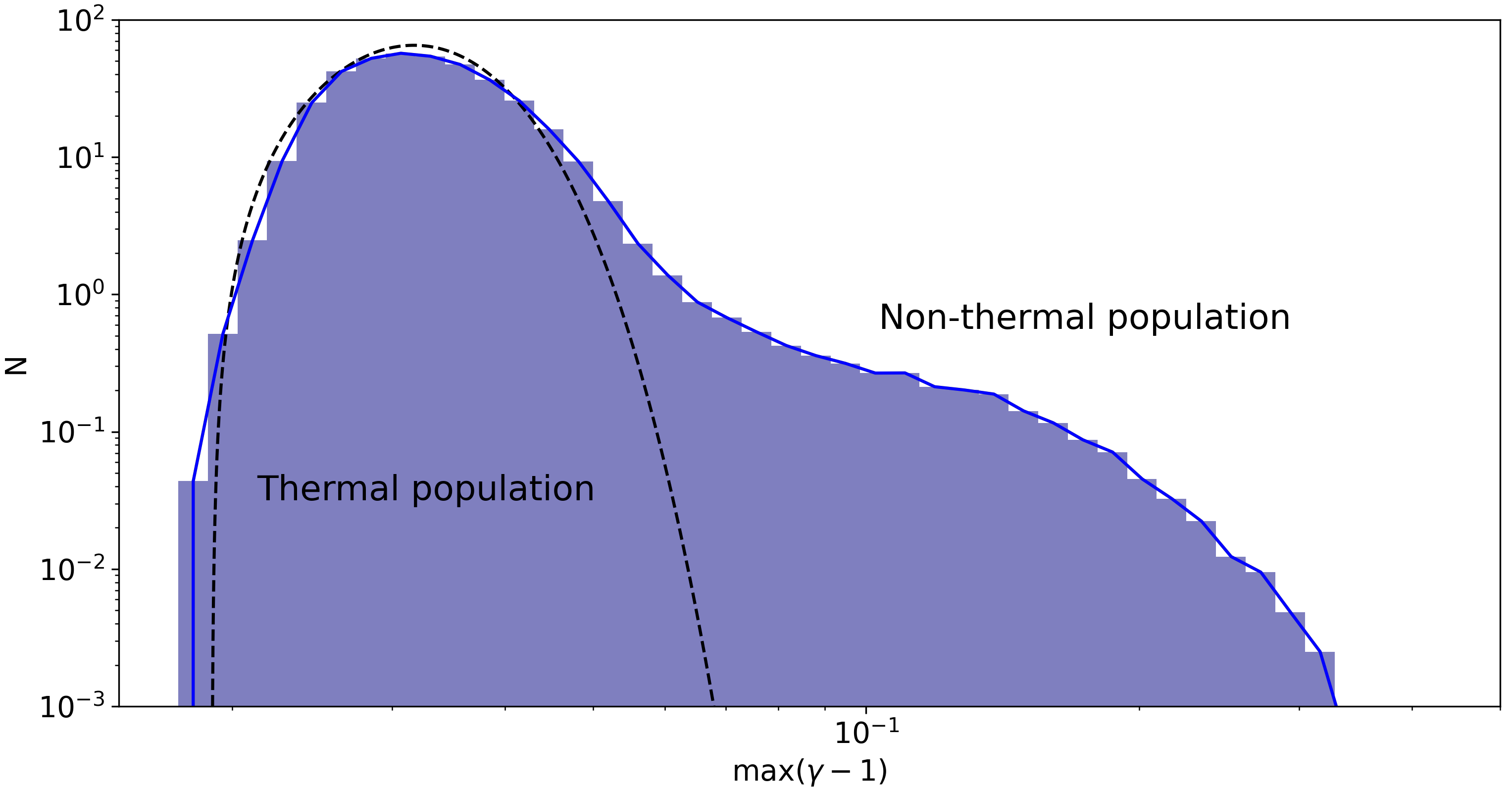}
 \caption{Histogram of the maximum kinetic energy of the particles in the dataset. The thermal electron population is represented by a dashed line for reference.
 }
 \label{fig:histogram}
\end{SCfigure}

The dataset we use in this work has been obtained from our recent PIC simulations of non-relativistic high Mach number shocks \cite{bohdan_2,bohdan_4}, in which the Buneman instability \cite{buneman} is excited in the shock foot. The dataset consists of $N_p=$ 252,000 electrons that pass through the Buneman instability region, with some of them undergoing energization.
We record these particle momenta, $\mathbf{p}$, as well as the electric field, $\mathbf{E}$, and magnetic field, $\mathbf{B}$, at their respective positions, across 1200 time steps. 
For the analysis presented in this paper we use only particle momentum data. For each particle we compute a label, $y_i=\max{(\gamma_i-1)}$ representing the maximum kinetic energy achieved by that particle along its trajectory, where $\gamma_i$ is the Lorentz factor of the $i$th particle. Figure \ref{fig:bun_traj} shows two examples of particle trajectories from our data.
The trajectory for a particle that is influenced by the Buneman instability (illustrated in panel (a) in a map of the $E_x$ electric field) and accelerated is shown by the red line, while the blue line depicts a particle that remains unaffected. This distinction is particularly evident in panel (b), which shows the evolution of the particle Lorentz factor. The shape of the momentum time series varies depending on whether the particle has undergone acceleration or not, and the neural network has the capability to discern this difference. This can be seen in panels (d), (e) and (f) for the two selected particles but in most cases the difference between energised and thermal particles is not so clear. 
Figure~\ref{fig:histogram} shows the histogram of the maximum kinetic energy. One can see that only a small fraction, approximately 2\%, of the total particle population are affected by the instability. The majority of particles belong to the thermal population, whereas for the non-thermal population, the number of particles decreases significantly with the energy.

\section{Results}
\subsection{Convolutional Neural Networks}
Convolutional Neural Networks (CNNs) are a type of NN that use convolutional layers to process data. These layers consist of multiple filters or units. Each filter contains a convolutional window with a given width, filled with weights that perform convolutions across the time series. The resulting output is then standardized, which helps the NN in effectively processing the data. Without standardization, the performance of NN would deteriorate. Finally, the standardized data goes through an activation function, often Leaky ReLU \cite{lrelu}, that gives the NN the characteristic non-linear behaviour. Typically, NN consist of multiple convolutional layers, varying from 3-5 to even hundreds of layers.

In our study, we investigate the use of a classical CNN 
for regression analysis on our data. Additionally, we employ an autoencoder based on a CNN to perform anomaly detection.

\subsection{Regression}
\begin{figure}
    \centering
\includegraphics[width=0.8\linewidth]{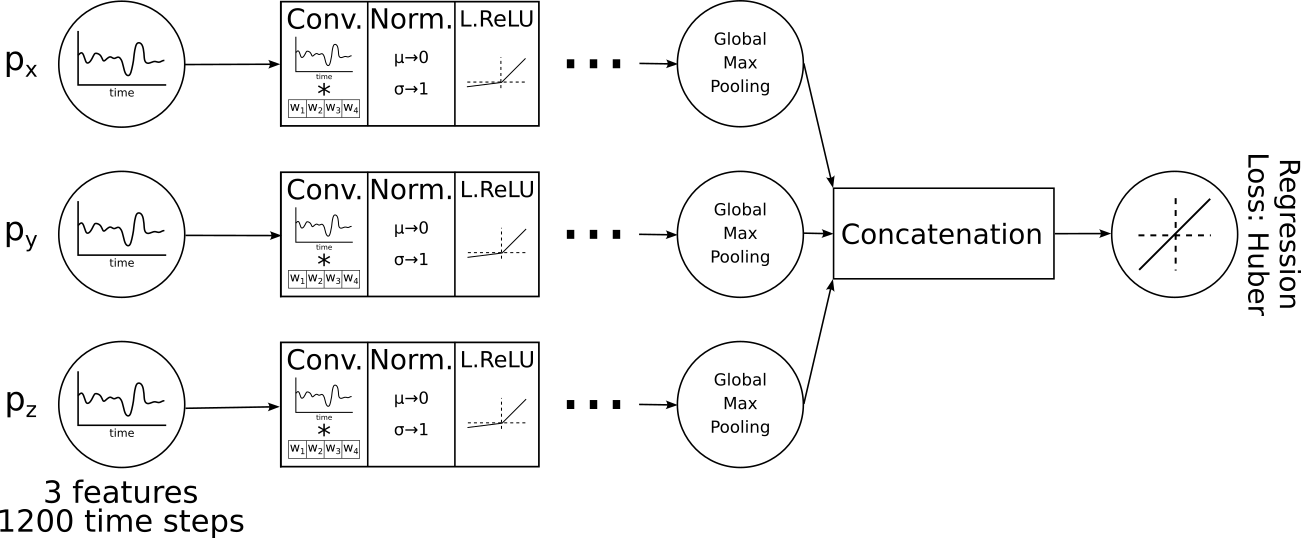}
    \caption{Sketch of a Convolutional Neural Network for a regression algorithm with input data representing the momentum time series.}
    \label{fig:cnn}
\end{figure}

The regression algorithm is a supervised learning method that provides a single value as its output. After the convolutional layers, the output passes through Max Pooling that selects the maximum value for each filter in the layer. The resulting data is concatenated into a single array and then it goes through the linear activation function, $y=x$, which predicts the maximum kinetic energy. This prediction is compared to the original value $y_i$ by using a loss function, specifically the Huber loss function~\cite{huber}, to update the weights of the convolutional layer filters (see Fig.~\ref{fig:cnn}). Due to the heavily unbalanced dataset, we use sample weights~\cite{sample_weights} on our input data, which helps balancing the input data.

\begin{figure}
    \centering
    \begin{subfigure}[c]{0.45\textwidth}
        \includegraphics[width=\linewidth]{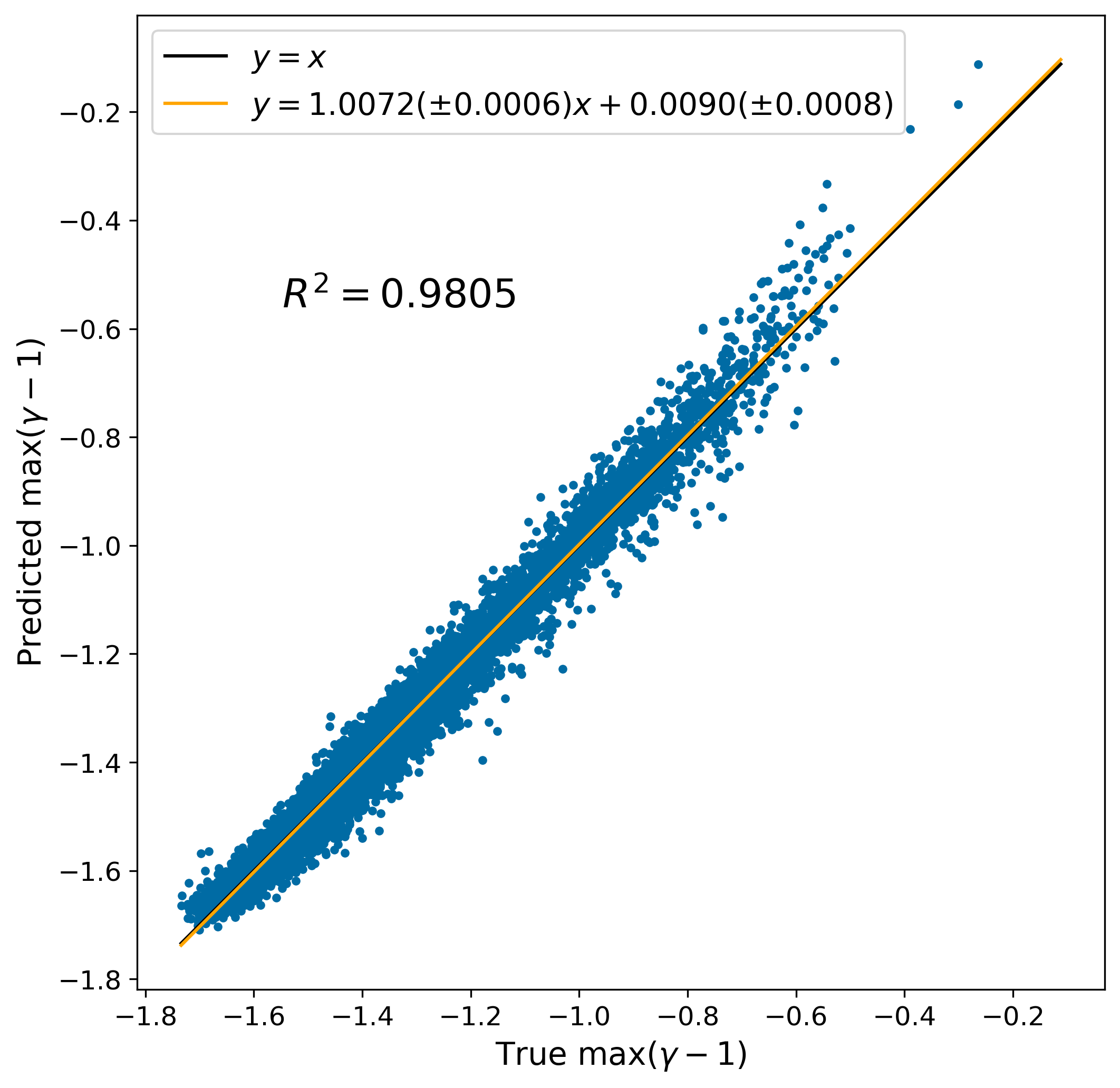}
    \end{subfigure}
    \begin{subfigure}[c]{0.54\textwidth}
        \includegraphics[width=\linewidth]{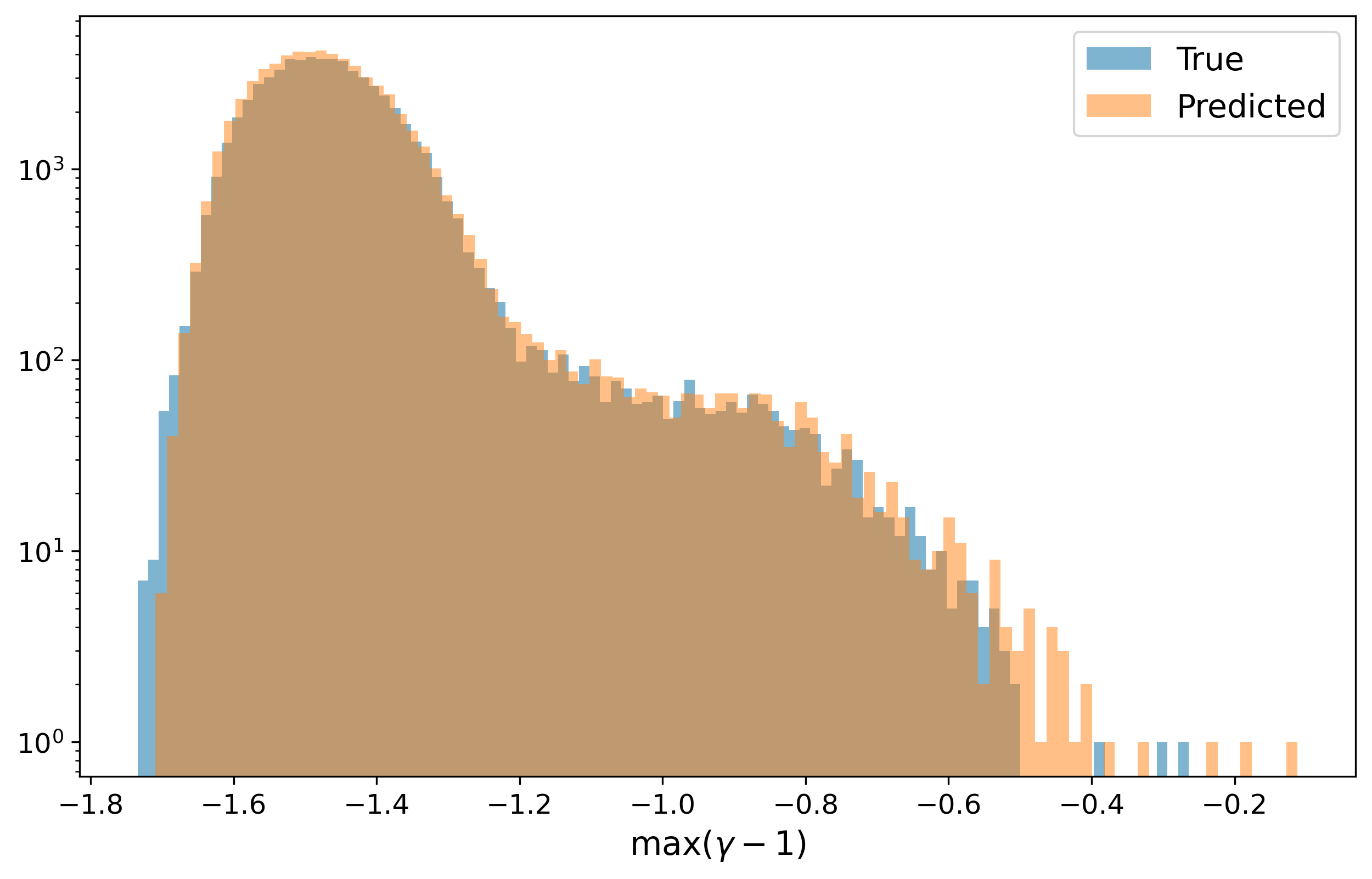}
    \end{subfigure}
    \caption{Results for regression. Left plot shows the comparison between the true value (x-axis) and predicted value (y-axis) of the maximum kinetic energy. Right plot shows the histograms for the true and predicted values of the maximum kinetic energy.}
    \label{fig:reg_results}
\end{figure}

Figure \ref{fig:reg_results} (left) shows the 
relation between the 
original and predicted values of $y_i$ in regression analysis. Our goal is to obtain a regression line close to $y=x$ and a high $R^2$ score. Remarkably, we achieve a score of 0.9805, despite a noisy momentum data. Figure \ref{fig:reg_results} (right) shows histograms comparing the true and predicted values of the maximum kinetic energy. Both histograms closely align, although there is a slight over-prediction in the energetic regime, possibly due to the impact of sample weighting in high-energy particles. 

\subsection{Anomaly Detection}
\begin{figure}
    \centering
    \includegraphics[width=0.75\linewidth]{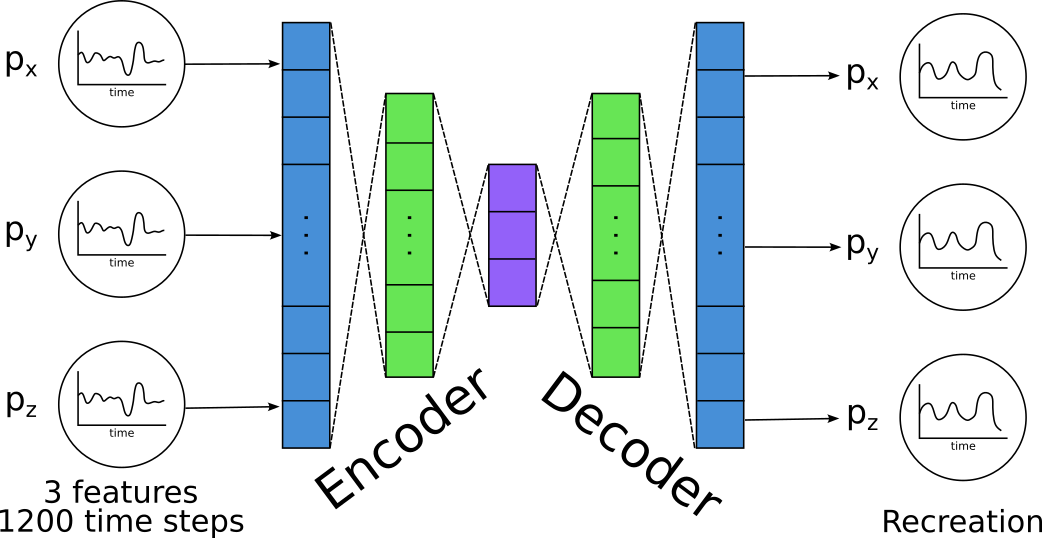}
    \caption{Sketch of an autoencoder used for anomaly detection. The data is encoded via several convolutional layers, culminating in a bottleneck layer, subsequently to be decoded to reconstruct the original data.}
    \label{fig:ad}
\end{figure}

Anomaly Detection is an unsupervised learning method that uses an autoencoder (Fig.~\ref{fig:ad}). The autoencoder operates as follows: The input data goes through several convolutional layers with progressively lower number of filters, effectively encoding the data.
This is very similar to the regression approach. However, the data is compressed into a bottleneck convolutional layer, containing fewer filters, but retaining essential information. Subsequently, the data is decoded, with the amount of filters in convolutional layers progressively increasing until a recreation of the original data is obtained. 
If the reconstructed time series significantly deviates from the original, it is tagged as an anomaly. 

To distinguish between anomalies and non-anomalies, a threshold is computed, equivalent to the maximum value achieved by the loss function (in our case, LogCosh). 
If a particle loss exceeds the threshold, it is labeled as an anomaly. It is important to note that we do not use any labels, as this is unsupervised learning. To train the autoencoder, we only use particles with $\max{(\gamma-1)}<0.07$ (non-energetic). This is
to ensure that the training phase only sees most commonly occurring particles.

\begin{figure}
    \centering
    \begin{subfigure}[c]{\textwidth}
        \makebox[\textwidth][c]{\includegraphics[width=1.0\textwidth]{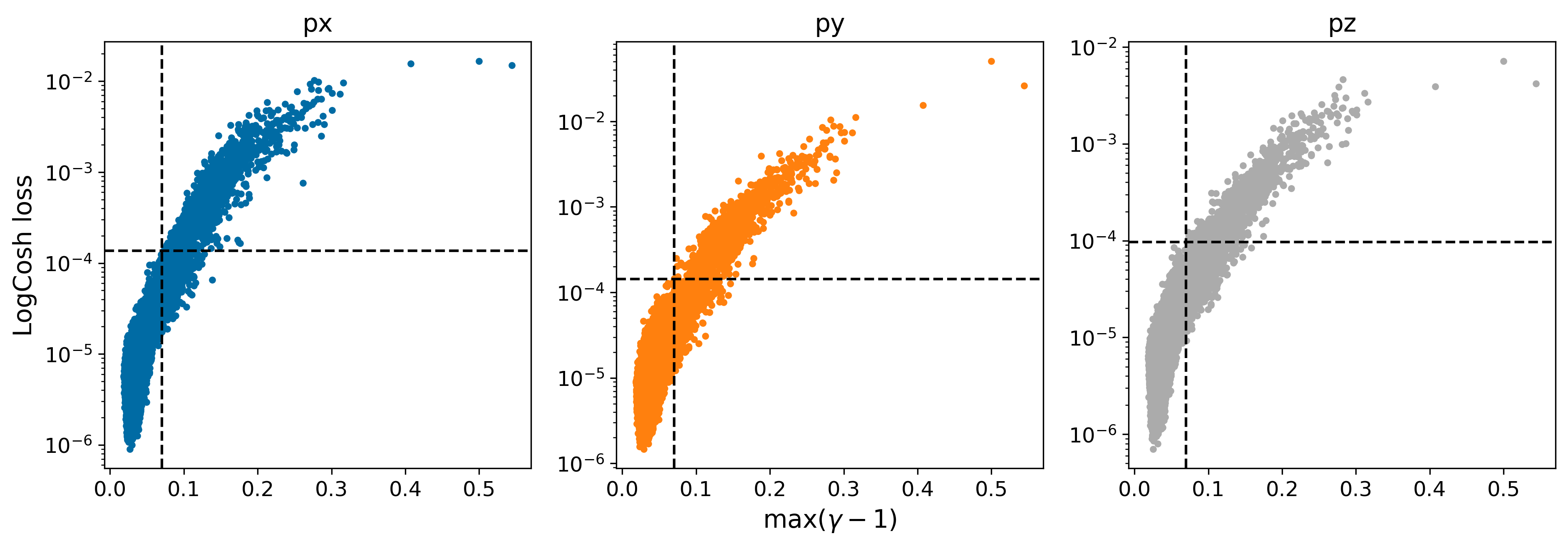}}
    \end{subfigure}
    \begin{subfigure}[c]{\textwidth}
        \centering
        \begin{tabular}{c|c|c}
                        & Non-energetic & Energetic \\ \hline
            Anomaly     &            17 &      1188 \\ \hline
            Non-anomaly &         61468 &       327
        \end{tabular}
    \end{subfigure}
    \caption{Results for Anomaly Detection. The top row shows the label of the particle vs. its corresponding testing loss, for the three features in the data, e.g., the particle momentum components. The horizontal dashed line represents the anomaly threshold, and the vertical one our threshold of $\max{(\gamma-1)}=0.07$ that separates the} energetic particles from the non-energetic ones. The table at the bottom shows the values that are in the four quadrants in the top plots. Particles that have at least one feature (one momentum component) with a higher loss than the threshold are tagged as anomaly.
    \label{fig:ad_results}
\end{figure}

Figure \ref{fig:ad_results} (top) shows the relation between the maximum kinetic energy of particles in the dataset and the loss obtained from the autoencoder. The horizontal dashed line represents the threshold for each input value (feature), whereas the vertical line represents our separation between energetic and non-energetic particles. 
Ideally, all energetic particles would reside in the top-right quadrant. However, near the crossing point it is harder for the NN  to differentiate between energised and non-energised particles, as their time series are relatively similar. Figure \ref{fig:ad_results} (top) can be translated into the table (Fig.~\ref{fig:ad_results}, bottom), which displays the number of energetic and non-energetic particles along with their assigned tags, either anomaly or non-anomaly. Overall, the autoencoder demonstrates an accurate prediction of energised particles just based on the shape of the time series, although 327 particles ($\sim$22\%) are not correctly identified. Conversely, for non-energetic particles, the autoencoder effectively distinguishes them, as evidenced by the low count of 17 compared to 61,468.

\begin{figure}
    \centering
    \begin{subfigure}{\textwidth}
        \makebox[\textwidth][c]{\includegraphics[width=1.\textwidth]{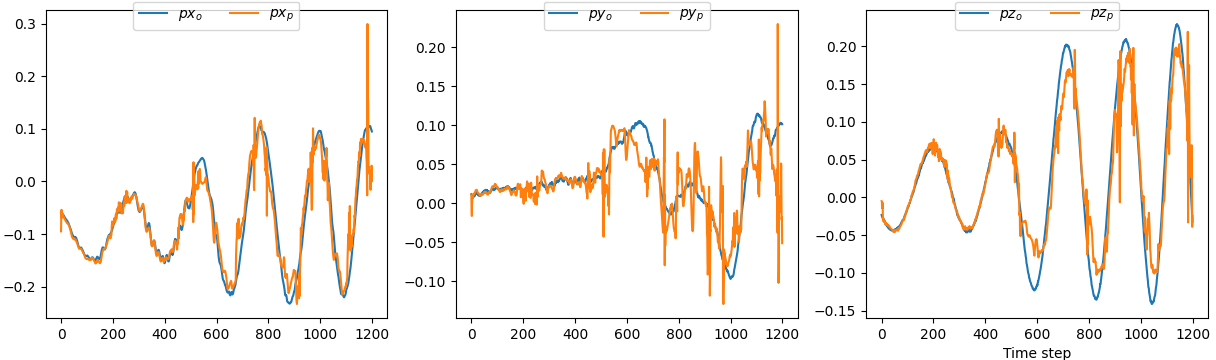}}
        \caption{Anomaly}
        \label{fig:cnn_ad_p_anomaly}
    \end{subfigure}
    \begin{subfigure}{\textwidth}
         \makebox[\textwidth][c]{\includegraphics[width=1.\textwidth]{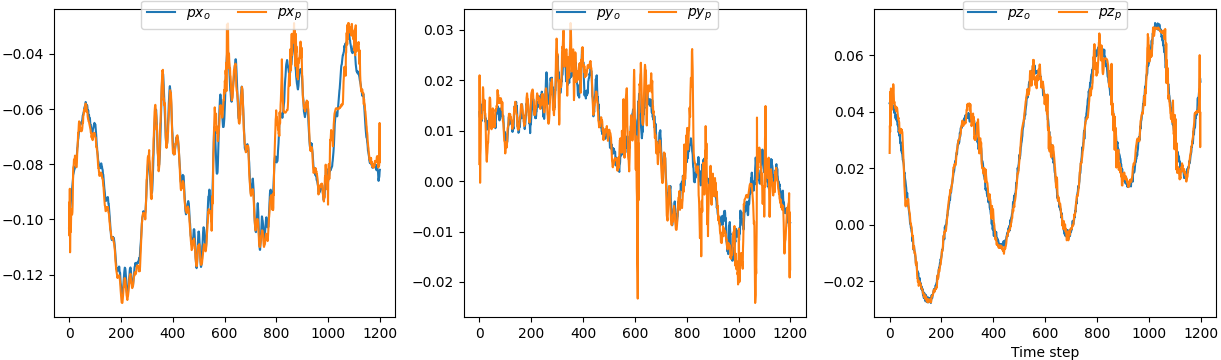}}
        \caption{Non-anomaly}
        \label{fig:cnn_ad_p_nonanomaly}
    \end{subfigure}
    \caption{Examples of how the Anomaly Detector re-creates the input data for each component of particle momentum. The blue line shows the original time series, while the orange line the predicted time series. If the recreation significantly differs from the input (high loss, top) the particle is tagged as an anomaly. Conversely, if the time series are similar, the particle is classified as a non-anomaly (low loss, bottom).}
    \label{fig:cnn_ad_p_examples}
\end{figure}

Figure \ref{fig:cnn_ad_p_examples} illustrates the predictions made by the NN on the data. In the case of an anomaly, the NN fails to recreate one of the time series and therefore it is tagged as an anomaly. In the example shown in the top row, the $p_x$ and $p_z$ data are satisfactorily reproduced. However, $p_y$ data deviate at an instance near 600-750 time step, marking the operation of the Buneman instability that affects $p_y$ as the particles surf the electrostatic waves and get energised. On the other hand, the non-anomaly is accurately recreated, and although there are slight deviations in some instances, the loss is below the anomaly threshold.


\section{Conclusions}
We employ NN to predict the maximum kinetic energy of particles that are energised through surfing on the Buneman instability waves in the foot of a quasi-perpendicular shock. Regression yields an excellent result, with a $R^2$ score of 0.9805. The anomaly detection 
allows us to detect energetic particles without requiring any labels, contrarily to regression which uses labels. Using only the momentum time series as input data, the autoencoder identifies around 78\% of energetic particles.
By incorporating this autoencoder in the analysis, the NN can aid in identifying energetic particles.

The results demonstrate our ability to predict the particle energy solely using momenta in a specific scenario of acceleration at the Buneman instability waves. However, we are also exploring the use of the electric field data as input to predict the maximum kinetic energy. Unlike momentum, the electric field at particle location is not directly related to the energy. The goal of this approach
is to test various acceleration scenarios beyond the surfing on the Buneman instability waves and experiment with other input variables, such as the Fourier transform of the input data. As simulations grow in size, tools such as NN facilitate faster and more precise analysis, enabling improved workflow in particle acceleration research.

\section*{Acknowledgments}
\noindent
The work of G.~T.~P. and J.~N. has been supported by Narodowe Centrum Nauki through research project no. 2019/33/B/ST9/02569. A.~B. was supported by the German Research Foundation (DFG) as part of the Excellence Strategy of the federal and state governments - EXC 2094 - 390783311. We gratefully acknowledge Polish high-performance computing infrastructure PLGrid (HPC Centers: ACK Cyfronet AGH) for providing computer facilities and support within computational grant no. PLG/2023/016378. This research was supported by the International
Space Science Institute (ISSI) in Bern, through ISSI International
Team project \#520 (Energy Partition Across Collisionless Shocks).

\end{document}